\documentclass[10pt,pra,twocolumn,amssymb,nofootinbib,showpacs]{revtex4}
\usepackage{bm}
\usepackage{graphicx}
\usepackage{amsmath}

\begin{document}
\title{\bf Collapse models: analysis of the free particle dynamics}
\author{Angelo Bassi}
\email{bassi@mathematik.uni-muenchen.de} \affiliation{Mathematisches
Institut der Universit\"at M\"unchen, Theresienstr. 39, 80333
M\"unchen, Germany, \\ The Abdus Salam International Centre for
Theoretical Physics, Strada Costiera 11, 34014 Trieste, Italy.}
\begin{abstract}
We study a model of spontaneous wavefunction collapse for a free
quantum particle. We analyze in detail the time evolution of the
single--Gaussian solution and the double--Gaussian solution,
showing how the reduction mechanism induces the localization of
the wavefunction in space; we also study the asymptotic behavior
of the general solution. With an appropriate choice for the
parameter $\lambda$ which sets the strength of the collapse
mechanism, we prove that: i) the effects of the reducing terms on
the dynamics of microscopic systems are negligible, the physical
predictions of the model being very close to those of standard
quantum mechanics; ii) at the macroscopic scale, the model
reproduces classical mechanics: the wavefunction of the center of
mass of a macro--object behaves, with high accuracy, like a point
moving in space according to Newton's laws.
\end{abstract}
 \maketitle

\section{Introduction}

Models of spontaneous wavefunction collapse
\cite{grw1,grw2,grw6,pp1,di1,di3, bel1,bel2,bla1,jp1,jp2,ip,
ad1,ad2,bar1} have reached significant results in providing a
solution to the measurement problem of quantum mechanics. This
goal is achieved by modifying the Schr\"odinger equation, adding
appropriate non--linear stochastic terms\footnote{If one aims at
reproducing the process of wavefunction collapse in
measurement--like situation, the terms which have to be added to
the Schr\"odinger equation must be non--linear and stochastic,
since these two are the characteristic features of the quantum
collapse process.}: such terms do not modify appreciably the
standard quantum dynamics of microscopic systems; at the same
time, they rapidly reduce the superposition of two or more
macroscopically different states of a macro--object into one of
them; in particular, they guarantee that measurements made on
microscopic systems always have definite outcomes, and with the
correct quantum probabilities. In this way, collapse models
describe --- within one single dynamical framework --- both the
quantum properties of microscopic systems and the classical
properties of macroscopic objects, providing a unified description
of micro-- and macro--phenomena.

In this paper, we investigate the physical properties of a
collapse model describing the (one--dimensional) evolution of a
free quantum particle subject to spontaneous localizations in
space; its dynamics is governed by the following stochastic
differential equation in the Hilbert space $L^2({\mathbb R})$:
\begin{eqnarray} \label{nle}
d\,\psi_{t}(x) & = & \left[ -\frac{i}{\hbar}\, \frac{p^2}{2m}\, dt
+ \sqrt{\lambda}\, (q - \langle
q \rangle_{t})\, dW_{t}\right. \nonumber \\
& & \left. - \frac{\lambda}{2}\, (q - \langle q \rangle_{t})^2 dt
\right] \psi_{t}(x),
\end{eqnarray}
where $q$ and $p$ are the position and momentum operators,
respectively, and $\langle q \rangle_{t} \equiv \langle \psi_{t} | q
| \psi_{t} \rangle$ denotes the quantum average of the operator $q$;
$m$ is the mass of the particle, while $\lambda$ is a positive
constant which sets the strength of the collapse mechanism. The
stochastic dynamics is governed by a standard Wiener process
$W_{t}$, defined on the probability space $(\Omega, {\mathcal F},
{\mathbb P})$ with the natural filtration $\{ {\mathcal F}_{t}: t
\geq 0 \}$ defined on it.

The value of the collapse constant $\lambda$ is given by the
formula\footnote{Differently from \cite{di3}, we assume that
$\lambda$ is proportional to the mass of the particle.}:
\begin{equation}
\lambda \; = \; \frac{m}{m_{0}}\, \lambda_{0},
\end{equation}
where $m_{0}$ is a reference mass which we choose to be equal to
that of a nucleon, and $\lambda_{0}$ is a fixed constant which we
take equal to:
\begin{equation}
\lambda_{0} \; \simeq \; 10^{-2} \, \makebox{m$^{-2}$ sec$^{-1}$};
\end{equation}
this value corresponds to the product of the two parameters
$\Lambda_{\makebox{\tiny GRW}}$ and $\alpha_{\makebox{\tiny GRW}}$
of the GRW collapse--model \cite{grw1}, where
$\Lambda_{\makebox{\tiny GRW}} \simeq 10^{-16}$ sec$^{-1}$ is the
localization rate for a nucleon and
$1/\sqrt{\alpha_{\makebox{\tiny GRW}}} \simeq 10^{-7}$ m is the
width of the Gaussian wavefunction inducing the localizations.

Eq. (\ref{nle}) has already been subject to investigation: in ref
\cite{jp1} a theorem proves the existence and uniqueness of strong
solutions\footnote{Existence and uniqueness theorems for a wide
class of stochastic differential equations whose coefficients are
{\it bounded} operators, are studied in ref \cite{bar3}. The case
of {\it unbounded} operators is covered in ref \cite{hol1}. See
ref \cite{pra, barr} for an introduction to stochastic
differential equation in infinite dimensional spaces.}; in refs
\cite{jp1,di1,bel2,hal} some properties of the solutions have been
analyzed: in particular, it has been shown that Gaussian
wavefunctions are solutions of Eq. (\ref{nle}), that their spread
reaches an asymptotic finite value (we speak in this case of a
``stationary'' solution), and that the general solution reaches
asymptotically a stationary Gaussian solution. Finally, in ref
\cite{di3}, Eq. (\ref{nle}) has been first proposed as a universal
model of wavefunction collapse.

The aim of our work is to provide a detailed analysis of the
physical properties of the solutions of Eq. (\ref{nle}). After
some mathematical preliminaries (Sec. \ref{math}), we will study
the time evolution of the two most interesting types of
wavefunctions: the single--Gaussian (Sec. \ref{1gauss}) and the
double--Gaussian wavefunctions (Sec. \ref{2gauss}); in Sec.
\ref{gs} we will discuss the asymptotic behavior of the general
solution.

We will next study the effects of the stochastic dynamics on
microscopic systems (Sec. \ref{micro}) and on macroscopic objects
(Sec. \ref{macro}); in the first case, we will see that the
prediction of the model are very close to those of standard
quantum mechanics, while in the second case we will show that the
wavefunction of a macro--object is well localized in space and
behaves like a point moving in space according to the classical
laws of motion. We end up with some concluding remarks (Sec.
\ref{concl}).

\section{Linear Vs Non Linear Equation} \label{math}

The easiest way to find solutions of a non--linear equation is ---
when feasible --- to linearize it: this is possible for Eq.
(\ref{nle}) and the procedure is well known in the literature
\cite{grw2,grw6,jp1,bar1,bar3,hol1}. Let us consider the following
linear stochastic differential equation:
\begin{equation} \label{le}
d\,\phi_{t}(x) \; = \; \left[ -\frac{i}{\hbar}\, \frac{p^2}{2m}\,
dt + \sqrt{\lambda}\, q \, d\xi_{t} - \frac{\lambda}{2}\, q^2 dt
\right] \phi_{t}(x);
\end{equation}
$\xi_{t}$ is a standard Wiener process defined on the probability
space $(\Omega, {\mathcal F}, {\mathbb Q})$, where ${\mathbb Q}$
is a new probability measure, whose connection with ${\mathbb P}$
will be clear in what follows. Contrary to Eq. (\ref{nle}), the
above equation does not preserve the norm of statevectors, so let
us define the normalized vectors:
\begin{equation} \label{nv}
\psi_{t} \; = \; \left\{
\begin{array}{cl}
\phi_{t}/\|\phi_{t}\| & \makebox{if: $\|\phi_{t}\| \neq 0$,} \\
\makebox{a fixed unit vector}\quad & \makebox{if: $\|\phi_{t}\| =
0$.}
\end{array}
\right.
\end{equation}
By using It\^o calculus, it is not difficult to show that
$\psi_{t}$ defined by (\ref{nv}) is a solution of Eq. (\ref{nle}),
whenever $\phi_{t}$ solves Eq. (\ref{le}). We now briefly explain
the relations between the two probability measures ${\mathbb Q}$
and ${\mathbb P}$, and between the two Wiener processes $\xi_{t}$
and $W_{t}$.

The key property of Eq. (\ref{le}) is that $p_{t} \equiv
\|\phi_{t}\|^2$ is a martingale \cite{jp1,hol1} satisfying the
following stochastic differential equation:
\begin{equation}
d p_{t} \; = \; 2\sqrt{\lambda}\, \langle q \rangle_{t}\, p_{t}\,
d\xi_{t},
\end{equation}
with $\langle q \rangle_{t} = \langle \psi_{t} | q | \psi_{t}
\rangle$. As a consequence of the martingale property (and assuming,
as we shall always do, that $\|\phi_{0}\| = 1$) $p_{t}$ can be used
to generate a new probability measure $\tilde{\mathbb P}$ on
$(\Omega, {\mathcal F})$ \cite{barr}; we choose ${\mathbb Q}$ in
such a way that the new measure $\tilde{\mathbb P}$ coincides with
${\mathbb P}$.

Given this, Girsanov's theorem \cite{ls} provides a simple relation
between the Wiener process $\xi_{t}$ defined on $(\Omega, {\mathcal
F}, {\mathbb Q})$, and the Wiener process $W_{t}$ defined on
$(\Omega, {\mathcal F}, {\mathbb P})$:
\begin{equation} \label{gt}
W_{t} \; = \; \xi_{t} - 2\sqrt{\lambda} \int_{0}^{t} \langle q
\rangle_{s}\, ds.
\end{equation}

The above results imply that one can find the solution $\psi_{t}$
of Eq. (\ref{nle}), given the initial condition $\psi_{0}$, by
using the following procedure:
\begin{enumerate}
\item Find the solution $\phi_{t}$ of Eq. (\ref{le}), with the initial
condition $\phi_{0} = \psi_{0}$.
\item Normalize the solution: $\phi_{t} \rightarrow \psi_{t} =
\phi_{t}/\| \phi_{t} \|$.
\item Make the substitution: $d\xi_{t} \rightarrow dW_{t} +
2 \sqrt{\lambda} \langle q \rangle_{t}$.
\end{enumerate}
The advantage of such an approach is that one can exploit the linear
character of Eq. (\ref{le}) to analyze the properties of the
nonlinear Eq. (\ref{nle}); as we shall see, the difficult part will
be computing $\langle q \rangle_{t}$ from $\phi_{t}$ (step 3): this
is where non linearity enters in a non--trivial way.

\section{Single Gaussian solution} \label{1gauss}

We start our analysis by taking, as a solution, a single--Gaussian
wavefunction\footnote{See ref \cite{pref} for an analogous
discussion within the CSL \cite{grw2} model of wavefunction
collapse.}:
\begin{equation} \label{gsol}
\phi_{t}^{\makebox{\tiny S}}(x) = \makebox{exp}\left[ - a_{t} (x -
\overline{x}_{t})^2 + i \overline{k}_{t}x + \gamma_{t}\right],
\end{equation}
where $a_{t}$ and $\gamma_{t}$ are supposed to be complex
functions of time, while $\overline{x}_{t}$ and $\overline{k}_{t}$
are taken to be real\footnote{For simplicity, we will assume in
the following that the initial values of these parameters do not
depend on $\omega \in \Omega$.}. By inserting (\ref{gsol}) into
Eq. (\ref{le}), one finds the following sets of stochastic
differential equations\footnote{The superscripts ``R'' and ``I''
denote, respectively, the real and imaginary parts of the
corresponding quantities.}:
\begin{eqnarray}
d a_{t} & = & \left[ \lambda - \frac{2i\hbar}{m}
\left(a_{t}\right)^2\right] dt \label{p1}\\
d \overline{x}_{t} & = &\frac{\hbar}{m}\, \overline{k}_{t}\, dt +
\frac{\sqrt{\lambda}}{2a_{t}^{\makebox{\tiny R}}} \left[
d{\xi}_{t} - 2\sqrt{\lambda}\overline{x}_{t}\, dt \right]
\label{p2} \\
d \overline{k}_{t} & = & - \sqrt{\lambda}\,
\frac{a_{t}^{\makebox{\tiny I}}}{a_{t}^{\makebox{\tiny R}}} \left[
d{\xi}_{t} - 2\sqrt{\lambda} \overline{x}_{t}\, dt \right]
\label{p3} \\
d\gamma_{t}^{\makebox{\tiny R}} & = & \left[ \lambda
\overline{x}^{2}_{t} + \frac{\hbar}{m} a_{t}^{\makebox{\tiny I}}
\right]dt + \sqrt{\lambda} \overline{x}_{t}\left[ d{\xi}_{t} -
2\sqrt{\lambda}\overline{x}_{t}\, dt \right] \label{p4} \\
d\gamma_{t}^{\makebox{\tiny I}} & = & \left[ -\frac{\hbar}{m}
a_{t}^{\makebox{\tiny R}} - \frac{\hbar}{2m} \overline{k}^{2}_{t}
\right] dt \nonumber\\
& & + \sqrt{\lambda}\, \frac{a_{t}^{\makebox{\tiny
I}}}{a_{t}^{\makebox{\tiny R}}}\, \overline{x}_{t}\left[
d{\xi}_{t} - 2\sqrt{\lambda}\overline{x}_{t}\, dt \right].
\label{p5}
\end{eqnarray}
For a single--Gaussian wavefunction, the two equations for
$\gamma_{t}$ can be omitted since the real part of $\gamma_{t}$ is
absorbed into the normalization factor, while the imaginary part
gives an irrelevant global fase.

The normalization procedure is trivial, and also the Girsanov
transformation (\ref{gt}) is easy since, for a Gaussian wavefunction
like (\ref{gsol}), one simply has $\langle q \rangle_{t} =
\overline{x}_{t}$. We then have the following set of stochastic
differential equations for the relevant parameters:
\begin{eqnarray}
d a_{t} & = & \left[ \lambda - \frac{2i\hbar}{m}\,
\left( a_{t} \right)^{2}\right] dt, \label{efa} \\
d \overline{x}_{t} & = &\frac{\hbar}{m}\, \overline{k}_{t}\, dt +
\frac{\sqrt{\lambda}}{2a_{t}^{\makebox{\tiny R}}} \, dW_{ t}, \\
d \overline{k}_{t} & = & - \sqrt{\lambda}\,\,
\frac{a_{t}^{\makebox{\tiny I}}}{a_{t}^{\makebox{\tiny R}}} \,
dW_{t};
\end{eqnarray}
throughout this section, we will discuss the physical implications
of these equations.

\subsection{The equation for $a_{t}$}

Eq. (\ref{efa}) for $a_{t}$ can be easily solved \cite{bel2,jp1}:
\begin{equation} \label{sola}
a_{t} \; = \; c \tanh \left[ b\, t + k \right],
\end{equation}
with:
\begin{eqnarray} \label{inval}
c & = & (1-i) \frac{1}{2} \sqrt{\frac{m\lambda}{\hbar}}, \quad b =
(1+i)
\sqrt{\frac{\hbar\lambda}{m}}, \nonumber \\
k & = & \tanh^{-1}\left[ \frac{a_{0}}{c} \right].
\end{eqnarray}
After some algebra, one obtains the following analytical
expressions for the real and the imaginary parts of $a_{t}$:
\begin{eqnarray}
a_{t}^{\makebox{\tiny R}} & = & \;\;\; \frac{\lambda}{\omega}\,\,
\frac{\sinh (\omega t + \varphi_{1}) + \sin (\omega t +
\varphi_{2})}{ \cosh (\omega t + \varphi_{1}) + \cos (\omega t +
\varphi_{2})}, \qquad{} \label{rpa}\\
a_{t}^{\makebox{\tiny I}} & = & -\frac{\lambda}{\omega}\,\,
\frac{\sinh (\omega t + \varphi_{1}) - \sin (\omega t +
\varphi_{2})}{ \cosh (\omega t + \varphi_{1}) + \cos (\omega t +
\varphi_{2})}, \quad \label{ipa}
\end{eqnarray}
where we have defined the frequency:
\begin{equation}
\omega \; = \; 2\,\sqrt{\frac{\hbar \lambda_{0}}{m_{0}}} \; \simeq
\; 10^{-5} \; \makebox{sec$^{-1}$},
\end{equation}
which does not depend on the mass of the particle. The two
parameters $\varphi_{1}$ and $\varphi_{2}$ are functions of the
initial condition: $\varphi_{1} = 2 k^{\makebox{\tiny R}}$,
$\varphi_{2} = 2 k^{\makebox{\tiny I}}$.

An important property of $a_{t}^{\makebox{\tiny R}}$ is
positivity:
\begin{equation} \label{pos}
a_{0}^{\makebox{\tiny R}} > 0 \quad \longrightarrow \quad
a_{t}^{\makebox{\tiny R}} > 0 \qquad \forall\; t > 0,
\end{equation}
which guarantees that an initially Gaussian wavefunction does not
diverge at any later time. To prove this, we first note that the
denominator of (\ref{rpa}) cannot be negative; if it is equal to
zero then also the numerator is zero: this discontinuity can be
removed by using expression (\ref{sola}) for $a_{t}$, which
--- according to the values (\ref{inval}) for $b$ and $k$ --- is
analytic for any $t$. It is then sufficient to show that the
numerator remains positive throughout time. Let us consider the
function $f(t) = \sinh (\omega t + \varphi_{1}) + \sin (\omega t +
\varphi_{2})$; we have that $f(0)
> 0$ and $f'(t) \geq 0$ for any $t$, which implies that $f(t) > 0$
for any $t$, as desired. Note that positivity of
$a_{t}^{\makebox{\tiny R}}$ matches with the fact that Eq.
(\ref{nle}) preserves the norm of statevectors.

\subsection{The spread in position and momentum}

The time evolution of the spread in position and momentum of the
Gaussian wavefunction (\ref{gsol}),
\begin{eqnarray}
\sigma_{q}(t) & = & \sqrt{\langle q^2 \rangle - \langle q \rangle^2
} \; = \; \frac{1}{2}\sqrt{\frac{1}{a_{t}^{\makebox{\tiny
R}}}}, \nonumber \\
\sigma_{p}(t) & = & \sqrt{\langle p^2 \rangle - \langle p \rangle^2
} \; = \; \hbar\,\sqrt{\frac{(a_{t}^{\makebox{\tiny R}})^2 +
(a_{t}^{\makebox{\tiny I}})^2}{a_{t}^{\makebox{\tiny R}}}},
\end{eqnarray}
is given by the following analytical expressions:
\begin{eqnarray}
\sigma_{q}(t) & = & \sqrt{\frac{\hbar}{m\omega}}\,
\sqrt{\frac{\cosh (\omega t + \varphi_{1}) + \cos (\omega t +
\varphi_{2})}{\sinh (\omega t + \varphi_{1}) + \sin (\omega t +
\varphi_{2})}}, \label{sx} \\
\sigma_{p}(t) & = & \sqrt{\frac{\hbar m\omega}{2}}\,
\sqrt{\frac{\cosh (\omega t + \varphi_{1}) - \cos (\omega t +
\varphi_{2})}{\sinh (\omega t + \varphi_{1}) + \sin (\omega t +
\varphi_{2})}}. \label{sp}\;\;\;\;\;
\end{eqnarray}
Fig. \ref{fig1} shows the different time dependence of the spread
in position, as given by the Schr\"odinger equation and by the
stochastic equation:
\begin{figure}
\begin{center}
{\includegraphics[scale=0.43]{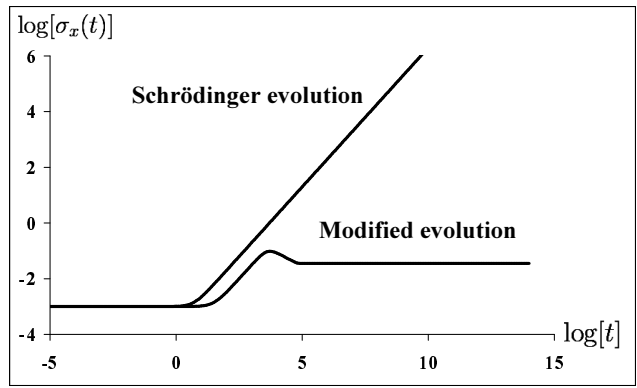}} \caption{The picture shows
the time evolution ($t$ measured in seconds) of the spread in
position $\sigma_{q}(t)$ (measured in meters) of a Gaussian
wavefunction, as given by the Schr\"odinger equation and by the
stochastic equation. The simulation has been made for a nucleon ($m
= m_{0}$). The initial spread has been taken equal to $10^{-3}$ m.}
\label{fig1}
\end{center}
\end{figure}
as we see, at the beginning the two evolutions almost coincide; as
time increases, while in the standard quantum case the spread goes
to infinity, the spread according to our stochastic equation reaches
the asymptotic value:
\begin{equation} \label{aval1}
\sigma_{q}(\infty) \; = \; \sqrt{\frac{\hbar}{m\omega}} \; \simeq
\; \left( 10^{-15} \sqrt{\frac{\makebox{Kg}}{m}}\, \right)\,
\makebox{m},
\end{equation}
which of course depends on the mass $m$ of the particle. This
behavior can be understood as follows: the reduction terms (which
tend to localize the wavefunction) and the standard quantum
Hamiltonian (which tends to spread out the wavefunction) compete
against each other until an equilibrium --- the stationary solution
--- is reached, which depends on the values of the parameters of the
model.

Also the spread in momentum changes in time, reaching the
asymptotic value:
\begin{equation} \label{aval2}
\sigma_{p}(\infty) \; = \; \sqrt{\frac{\hbar m\omega}{2}} \;
\simeq \; \left( 10^{-19} \sqrt{\frac{m}{\makebox{Kg}}}\,
\right)\, \frac{\makebox{Kg m}}{\makebox{sec}}
\end{equation}
It is interesting to compare the two asymptotic values for the
spread in position and momentum; one has:
\begin{equation}
\sigma_{q}(\infty)\, \sigma_{p}(\infty) \; = \;
\frac{\hbar}{\sqrt{2}}
\end{equation}
which corresponds to almost the minimum allowed by Heisenberg's
uncertainty relations \cite{grw1,hal}: the collapse model, then,
induces almost the best possible localization of the wavefunction
--- both in position and momentum. In accordance with \cite{di1},
any Gaussian wavefunction having these asymptotic values for
$\sigma_{q}$ and $\sigma_{p}$ will be called a ``stationary
solution''\footnote{Strictly speaking, a ``stationary'' solution is
not stationary at all, since both the mean in position and the mean
in momentum may change in time --- as it happens in our case; the
term ``stationary'' refers only to the spread of the wavefunction.}
of Eq. (\ref{nle}).

Note the interesting fact that the evolution of the spread in
position and momentum is {\it deterministic} and depends on the
noise only indirectly, through the constant $\lambda$.

\subsection{The mean in position and momentum}

The quantities $\langle q \rangle_{t} = \overline{x}_{t}$ and
$\langle p \rangle_{t} = \hbar \overline{k}_{t}$, corresponding to
the peak of the Gaussian wavefunction in the position and momentum
spaces, respectively, satisfy the following stochastic differential
equations:
\begin{eqnarray}
d \langle q \rangle_{t} & = & \frac{1}{m}\, \langle p \rangle_{t}
dt \; + \; \sqrt{\lambda}\, \frac{1}{2a_{t}^{\makebox{\tiny R}}}\,
dW_{t}, \\
d \langle p \rangle_{t} & = & - \sqrt{\lambda} \hbar\,
\frac{a_{t}^{\makebox{\tiny I}}}{a_{t}^{\makebox{\tiny R}}}\,
dW_{t}. \label{efp}
\end{eqnarray}
Their average values obey the {\it classical}
equations\footnote{These equations can be considered as the
stochastic extension of Ehrenfest's theorem.} for a free particle
of mass $m$:
\begin{eqnarray}
m\, \frac{d}{dt}\,{\mathbb  E}\left[ \langle q \rangle_{t} \right]
& = & {\mathbb  E}\left[ \langle p \rangle_{t} \right],
\label{mpa} \\
{\mathbb  E}\left[ \langle p \rangle_{t} \right] & = & \langle p
\rangle_{0}, \label{emim}
\end{eqnarray}
while the coefficients of covariance matrix
\begin{eqnarray}
C(t) & = & {\mathbb E} \left[ \left[
\begin{array}{c}
\langle q \rangle_{t} - {\mathbb E}[\langle q \rangle_{t}] \\
\langle p \rangle_{t} - {\mathbb E}[\langle p \rangle_{t}]
\end{array}
\right] \cdot \left[
\begin{array}{c}
\langle q \rangle_{t} - {\mathbb E}[\langle q \rangle_{t}] \\
\langle p \rangle_{t} - {\mathbb E}[\langle p \rangle_{t}]
\end{array}
\right]^{\top} \right] \nonumber \\
& & \equiv \left[
\begin{array}{cc}
C_{q^2}(t) & C_{qp}(t) \\
C_{pq}(t) & C_{p^2}(t)
\end{array}
\right] \nonumber
\end{eqnarray}
evolve as follows:
\begin{eqnarray}
\frac{d}{dt}\, C_{q^2}(t) & = & \frac{2}{m}\, C_{qp}(t) +
\frac{\lambda}{4} \frac{1}{(a_{t}^{\makebox{\tiny R}})^2}, \label{eq1c} \\
\frac{d}{dt}\, C_{qp}(t) & = & \frac{1}{m}\, C_{p^2}(t) -
\frac{\lambda \hbar}{2} \frac{a_{t}^{\makebox{\tiny
I}}}{(a_{t}^{\makebox{\tiny R}})^2},  \label{eq2c}  \\
\frac{d}{dt}\, C_{p^2}(t) & = & \lambda \hbar^2 \left(
\frac{a_{t}^{\makebox{\tiny I}}}{a_{t}^{\makebox{\tiny R}}}
\right)^2. \label{eq3c}
\end{eqnarray}
Particularly interesting is the third equation, which implies that
the wavefunction picks larger and larger components in momentum,
as time increases; as a consequence, the energy of the system
increases in time, as it can be seen by writing down the
stochastic differential for $\langle H \rangle_{t} \equiv \langle
p^2 \rangle_{t}/2m$:
\begin{equation}
d \langle H \rangle_{t} \; = \; \frac{\lambda \hbar^2}{2m}\, dt \;
- \; \sqrt{\lambda}\, \frac{\hbar}{m}\,
\frac{a_{t}^{\makebox{\tiny I}}}{a_{t}^{\makebox{\tiny R}}}\,
\langle p \rangle_{t}\, dW_{t},
\end{equation}
from which it follows that:
\begin{equation}
\frac{d}{dt}\, {\mathbb E}[\langle H \rangle_{t}] \; = \;
\frac{\lambda \hbar^2}{2m} \; = \; \frac{\lambda_{0}
\hbar^2}{2m_{0}} \; \simeq \; 10^{-43} \; \makebox{J/sec}.
\end{equation}
This energy non--conservation is a typical feature of
space--collapse models, but with our choice for the parameter
$\lambda$, the increase is so weak that it cannot be detected with
present--day technology (see ref. \cite{grw1}).

\section{Double Gaussian solution} \label{2gauss}

We now study the time evolution of the superposition of two
Gaussian wavefunctions; such an analysis is interesting since it
allows to understand in a quite simple and clear way how the
reduction mechanism works, i.e. how the superposition of two
different position states is reduced into one of them. To this
purpose, let us consider the following wavefunction:
\begin{eqnarray} \label{2gsol}
\phi_{t}^{\makebox{\tiny D}}(x) & = & \phi_{1t}(x) \, + \,
\phi_{2t}(x) \; = \; \nonumber \\
& = & \makebox{exp}\left[ - a_{1t} (x - \overline{x}_{1t})^2 + i
\overline{k}_{1t} x + \gamma_{1t}\right]
\nonumber \\
& + & \makebox{exp}\left[ - a_{2t} (x - \overline{x}_{2t})^2 + i
\overline{k}_{2t} x + \gamma_{2t}\right];
\end{eqnarray}
we follow the strategy outlined in Sec. II, by first finding the
solution of the linear equation.

Because of linearity, $\phi_{t}^{\makebox{\tiny D}}(x)$ is
automatically a solution of Eq. (\ref{le}), provided that its
parameters satisfy Eqs. (\ref{p1}) to (\ref{p5}). The difficult part
of the analysis is related to the change of measure: the reason is
that in the double--Gaussian case the quantum average $\langle q
\rangle_{t}$ is not simply equal to $\overline{x}_{1t}$ or
$\overline{x}_{2t}$, as it is for a single--Gaussian wavefunction,
but is a nontrivial function\footnote{This is the reason why the
dynamics of these parameters changes in a radical way, with respect
to the single--Gaussian case.} of all the parameters defining
$\phi_{t}^{\makebox{\tiny D}}(x)$; in spite of this difficulty, the
most interesting properties of the dynamical evolution of
$\phi_{t}^{\makebox{\tiny D}}(x)$ can be analyzed in a rigorous way.

We first observe that the two equations for $a_{1t}$ and $a_{2t}$
are deterministic and thus insensitive of the change of measure;
accordingly, the spread (both in position and in momentum) of the
two Gaussian functions defining $\phi_{t}^{\makebox{\tiny D}}(x)$
evolve {\it independently} of each other, and maintain all the
properties discussed in the previous section. For simplicity, we
assume that $a_{1t} = a_{2t}$ at $t = 0$ so that these two
parameters will remain equal at any subsequent time.

\subsection{The asymptotic behavior}
\label{twogaussas}

Let us consider the differences $X_{t} = \overline{x}_{2t} -
\overline{x}_{1t}$ and $K_{t} = \overline{k}_{2t} -
\overline{k}_{1t}$ between the peaks of the two Gaussian functions
in the position and in the momentum spaces; they satisfy the
following set of equations:
\begin{equation} \label{ls}
d \left[
\begin{array}{c}
X_{t} \\ K_{t}
\end{array}
\right] \; = \; \left[
\begin{array}{cc}
- A_{1}(t) & \hbar/m \\
- A_{2}(t) & 0
\end{array}
\right] \cdot \left[
\begin{array}{c}
X_{t} \\ K_{t}
\end{array}
\right] \, dt,
\end{equation}
with:
\begin{eqnarray}
A_{1}(t) & = & \;\omega\, \frac{\cosh (\omega t + \varphi_{1}) +
\cos (\omega t + \varphi_{2})}{\sinh (\omega t + \varphi_{1}) +
\sin (\omega t + \varphi_{2})}, \\
A_{2}(t) & = & 2\lambda\, \frac{\sinh (\omega t + \varphi_{1}) -
\sin (\omega t + \varphi_{2})}{\sinh (\omega t + \varphi_{1}) +
\sin (\omega t + \varphi_{2})}.
\end{eqnarray}
We see --- this is the reason why we have taken into account the
differences $X_t$ and $K_t$ --- that the above system of equations
is {\it deterministic}, so it does not depend on the change of
measure.

The coefficients of the $2\times 2$ matrix $A(t)$ defining the
linear system (\ref{ls}) are analytic in the variable $t$, and the
Liapunov's type numbers \cite{ltn} of the system  are the same as
those of the linear system obtained by replacing $A(t)$ with
$A(\infty)$, where:
\begin{equation}
A(\infty) \; \equiv \; \lim_{t \rightarrow + \infty} \, A(t) \; =
\; \left[
\begin{array}{cc}
-\omega & \hbar/m \\
-2\lambda & 0
\end{array}
\right].
\end{equation}
The eigenvalues of the matrix $A(\infty)$ are:
\begin{equation}
\mu_{1,2} \; = \; - \frac{1}{2}\,\left(1 \pm i\right)\,\omega,
\end{equation}
from which it follows that the linear system (\ref{ls}) has only
one Lyapunov's type number: $-\omega/2$. This implies that, for
any non trivial (vector) solution ${\bf Z}(t)$ of (\ref{ls}), one
has:
\begin{equation} \label{asbev}
\lim_{t \rightarrow + \infty}\, \frac{\ln |{\bf Z}(t)|}{t} \; = \;
- \frac{\omega}{2} \quad \Rightarrow \quad \lim_{t \rightarrow +
\infty}\, {\bf Z}(t) \; = \; {\bf 0}.
\end{equation}
We arrive at the following result: asymptotically, the difference
$\overline{x}_{2t} - \overline{x}_{1t}$ between the peaks of the
two Gaussian wavefunctions in the position space goes to zero;
also the difference $\overline{k}_{2t} - \overline{k}_{1t}$
between the peaks of the two Gaussian wavefunctions in the
momentum space vanishes. In other words, the two Gaussian
wavefunctions converge toward each other, and asymptotically they
become one single--Gaussian wavefunction which, from the analysis
of the previous subsection, is a ``stationary'' solution of the
stochastic equation (\ref{nle}).

Anyway, this behavior in general {\it cannot} be responsible for
the collapse of a macroscopic superposition; as a matter of fact,
let us consider the following situation which, for later
convenience, we call ``situation ${\mathcal A}$'':

\noindent i) The wavefunction is in a superposition of the form
(\ref{2gsol}) and, at time $t = 0$ (consequently, also for any later
$t$), $a_{1t}$ and $a_{2t}$ are equal to their asymptotic value;

\noindent ii) The distance  $K_{t} \equiv \overline{k}_{2t} -
\overline{k}_{1t}$ is zero at time $t = 0$.

\noindent Under these assumptions, the linear system (\ref{ls})
can be easily solved and one gets:
\begin{equation} \label{sita}
X_{t} \; = \; X_{0}\, e^{\displaystyle -\omega t/2}\, \left[ \cos
\frac{\omega t}{2} - \sin \frac{\omega t}{2} \right].
\end{equation}
We see that the time evolution of $X_{t}$ is independent of the
mass of the particle: this result implies that, when the spread of
the two Gaussian wavefunctions is equal to its asymptotic
value\footnote{We will see in Sec. \ref{macro} that at the
macroscopic level the spread of a Gaussian wavefunction converges
very rapidly towards its asymptotic value.}, their distance
decreases with a rate $\omega/2 \sim 10^{-5}$ sec$^{-1}$, which is
too slow to justify a possible collapse, in particular at the
macro--level.

\subsection{The collapse}

Now we show that the collapse of the wavefunction occurs because,
during the evolution, one of the two Gaussian wavefunctions is
suppressed with respect to the other one\footnote{Accordingly, the
collapse mechanism is precisely the same as the one of the
original GRW model \cite{grw1}.}; the quantity which measure this
damping is the difference $\Gamma_{t}^{\makebox{\tiny R}} =
\gamma_{2t}^{\makebox{\tiny R}} - \gamma_{1t}^{\makebox{\tiny
R}}$, which satisfies the stochastic differential equation:
\begin{equation} \label{eqfg}
d\Gamma_{t}^{\makebox{\tiny R}} =  - \lambda X_{t} \left[
\overline{x}_{1t} + \overline{x}_{2t} - 2\langle q \rangle_{t}
\right] dt + \sqrt{\lambda}X_{t} dW_{t};
\end{equation}
if $\Gamma_{t}^{\makebox{\tiny R}} \rightarrow +\infty$, then
$\phi_{1}$ is suppressed with respect to $\phi_{2}$ and the
superposition practically reduces to $\phi_{2}$; the opposite
happens if $\Gamma_{t}^{\makebox{\tiny R}} \rightarrow -\infty$.

To be more precise, we introduce a positive constant $b$ which we
assume to be conveniently large (let us say, $b = 10$) and we say
that the superposition is {\it suppressed} when
$|\Gamma^{\makebox{\tiny R}}| \geq b$; moreover, we say that:
\begin{eqnarray}
\makebox{$\phi^{\makebox{\tiny D}}$ is reduced to $\phi_{1}$
when:} &
& \Gamma^{\makebox{\tiny R}} \leq - b, \nonumber \\
\makebox{$\phi^{\makebox{\tiny D}}$ is reduced to $\phi_{2}$
when:} & & \Gamma^{\makebox{\tiny R}} \geq + b. \nonumber
\end{eqnarray}
We now study the time evolution of $\Gamma_{t}^{\makebox{\tiny
R}}$.

By writing $\langle q \rangle_{t}$ in terms of the coefficients
defining $\phi_{t}^{\makebox{\tiny D}}(x)$:
\begin{equation}
\langle q \rangle_{t} \; = \; \sqrt{\frac{\pi}{2
a_{t}^{\makebox{\tiny R}}}}\, \left[ \overline{x}_{1t} e^{2
\gamma_{1t}^{\makebox{\tiny R}}} + \overline{x}_{2t} e^{2
\gamma_{2t}^{\makebox{\tiny R}}} + \delta_{t}
e^{\gamma_{1t}^{\makebox{\tiny R}} + \gamma_{2t}^{\makebox{\tiny
R}}} \right]
\end{equation}
it is not difficult to prove that Eq. (\ref{eqfg}) becomes:
\begin{equation} \label{eqfg2}
d\Gamma_{t}^{\makebox{\tiny R}} = \lambda X_{t}^2 \, \tanh
\Gamma_{t}^{\makebox{\tiny R}} \, dt \, + \,
g_{t}(\Gamma_{t}^{\makebox{\tiny R}})\,dt \, + \,
\sqrt{\lambda}X_{t}\, dW_{t},
\end{equation}
where we have defined the following quantities:
\begin{eqnarray} \label{efg}
\delta_{t} & = & h_{t} \left[ (\overline{x}_{1t} +
\overline{x}_{2t}) \cos \theta_{t} + Y_{t} \sin \theta_{t} \right],
\\
g_{t}(\Gamma_{t}^{\makebox{\tiny R}}) & = & 2 \lambda X_{t} h_t
e^{\Gamma_{t}^{\makebox{\tiny R}}} \cdot \\
& & \frac{Y_{t} \sin \theta_{t} \left[ 1 +
e^{2\Gamma_{t}^{\makebox{\tiny R}}} \right] + X_{t} \cos
\theta_{t} \left[ 1 - e^{2\Gamma_{t}^{\makebox{\tiny R}}} \right]
}{\left[ 1 + e^{2\Gamma_{t}^{\makebox{\tiny R}}} + 2 h_t \cos
\theta_{t} e^{\Gamma_{t}^{\makebox{\tiny R}}} \right] \left[ 1 +
e^{2\Gamma_{t}^{\makebox{\tiny R}}} \right]} \nonumber
\end{eqnarray}
and:
\begin{eqnarray}
h_t & = & \exp\left[ -\frac{a_{t}^{\makebox{\tiny R}}}{2} \left(
X^{2}_{t} + Y_{t}^2 \right) \right], \\
Y_{t} & = & -\frac{2 a_{t}^{\makebox{\tiny I}} X_{t} +
K_{t}}{2a_{t}^{\makebox{\tiny R}}}, \\
\theta_{t} & = & \frac{1}{2}\, \left(\overline{x}_{1t} +
\overline{x}_{2t} \right) K_{t} + \Gamma_{t}^{\makebox{\tiny I}},
\qquad \Gamma_{t}^{\makebox{\tiny I}} =
\gamma_{2t}^{\makebox{\tiny I}} - \gamma_{1t}^{\makebox{\tiny I}},
\qquad
\end{eqnarray}
with $a_{t}^{\makebox{\tiny R}} \equiv a_{1t}^{\makebox{\tiny R}}
= a_{2t}^{\makebox{\tiny R}}$.

Eq. (\ref{eqfg2}) cannot be solved exactly, due to the presence of
the term $g_{t}(\Gamma_{t}^{\makebox{\tiny R}})$ which is a
non--simple function of $\Gamma_{t}^{\makebox{\tiny R}}$; to
circumvent this problem, we proceed as follows. We study the
following stochastic differential equation:
\begin{equation} \label{eqfg3}
d{\tilde \Gamma}_{t}^{\makebox{\tiny R}} = \lambda X_{t}^2 \,
\tanh {\tilde \Gamma}_{t}^{\makebox{\tiny R}} \, dt \, + \,
\sqrt{\lambda}X_{t}\, dW_{t},
\end{equation}
which corresponds to Eq. (\ref{eqfg2}) without the term
$g_{t}(\Gamma_{t}^{\makebox{\tiny R}})$, and at the end of the
subsection we estimate the error made by ignoring such a term.

In studying Eq. (\ref{eqfg3}), it is convenient to introduce the
following time--change:
\begin{equation} \label{tch}
t \; \longrightarrow \; s_{t} \; = \; \lambda \int_{0}^t X_{u}^2
\, du.
\end{equation}
$X_{t}^2$ is a continuous, differentiable and non--negative
function, which we can assume not to be identically
zero\footnote{If there exists an interval $I$ of ${\mathbb R}^+$
such that $X_{t} \equiv 0$ $ \forall t \, \in \, I$, then Eq.
(\ref{ls}) implies that $X_{t}$ remains equal to 0 for any
subsequent time, i.e. the two Gaussian wavefunctions coincide.} in
any sub--interval of ${\mathbb R}^+$; as a consequence, $s_{t}$ is
a monotone increasing --- thus invertible --- function of $t$ and
Eq. (\ref{tch}) defines a good time change.

Under this time substitution, Eq. (\ref{eqfg3}) becomes:
\begin{equation} \label{eqfg4}
d{\tilde \Gamma}_{s}^{\makebox{\tiny R}} = \tanh {\tilde
\Gamma}_{s}^{\makebox{\tiny R}} \, ds \, + \, d{\tilde W}_{s},
\end{equation}
where
\begin{equation} \label{nwp}
{\tilde W}_{s} \; = \; \sqrt{\lambda}\, \int_{0}^{s} X_{t} \, d
W_{t}
\end{equation}
is a Wiener process\footnote{See pages 111--113 of \cite{gsb}.} with
respect to the filtered space $( \Omega, {\mathcal F}, \{ {\mathcal
F}_{s}: s \geq 0 \}, {\mathbb P})$. Note that, since according to
Eq. (\ref{asbev}), $X_{t}$ in general decays exponentially in time,
$s_{\infty} < \infty$, and Eq. (\ref{eqfg4}) is physically
meaningful only within the interval\footnote{Thus, strictly
speaking, Eq. (\ref{nwp}) defines a Wiener process only for $s \leq
s_{\infty}$; we then extend in a standard way the process to the
interval $(s_{\infty}, \infty)$.} $[0, s_{\infty})$.

Eq. (\ref{eqfg4}) can be analyzed in great detail \cite{gsb}; in
particular, the following properties can be proven to
hold\footnote{Throughout the analysis, we will assume that
${\tilde \Gamma}_{0}^{\makebox{\tiny R}} \equiv b_{0} \, \in \,
(-b, b)$.}:

\noindent 1. Let us define the {\it collapse time} $S_{b} \equiv
\inf \{ s : |{\tilde \Gamma}_{s}^{\makebox{\tiny R}}| \, \geq \, b
\}$: this time is finite with probability 1 and its {\it average
value} is equal to\footnote{See theorem 2, page 108 of
\cite{gsb}.}:
\begin{equation} \label{ats}
{\mathbb E}[S_{b}] \; = \; b \tanh b \, - \, b_{0} \tanh b_{0}.
\end{equation}
If $b_{0} \ll b$ (which occurs when both terms of the
superposition give a non--vanishing contribution), and since we
have assumed $b \simeq 10$, then ${\mathbb E}[S_{b}] \simeq b$.

\noindent 2. The {\it variance} ${\mathbb V} [S_{b}] \equiv
{\mathbb E}[S_{b}^2] - {\mathbb E}^2[ S_{b} ]$ is given
by\footnote{See theorem 3, page 109 of \cite{gsb}.}:
\begin{equation}
{\mathbb V} [S_{b}] \; = \; F(b) \, - \, F(b_{0}),
\end{equation}
with:
\begin{equation}
F(x) = x^2 \tanh^2 x + x \tanh x - x^2.
\end{equation}
Note that since $F(x)$ is an even, positive function, increasing
for positive values of $x$, it follows that ${\mathbb V} [S_{b}] >
0$ whenever $|b_{0}| < b$, as it assumed to be.

\noindent 3. The {\it collapse probability} $P_{\phi_{2}}$ that
$\phi^{\makebox{\tiny D}}$ is reduced to $\phi_{2}$, i.e. that
${\tilde \Gamma}_{s}^{\makebox{\tiny R}}$ hits point $b$ before
point $-b$ is given by\footnote{See theorem 4, page 110 of
\cite{gsb}.}:
\begin{equation}
P_{\phi_{2}} \; = \; \frac{1}{2}\, \frac{\tanh b + \tanh
b_{0}}{\tanh b};
\end{equation}
according to our choice for $b$, $\tanh b \simeq 1$, and
consequently
\begin{eqnarray}
P_{\phi_{2}} & \simeq & \frac{1}{2}\, [ 1 + \tanh b_{0} ] \; = \;
\frac{e^{2\gamma_{20}^{\makebox{\tiny
R}}}}{e^{2\gamma_{10}^{\makebox{\tiny R}}} +
e^{2\gamma_{20}^{\makebox{\tiny R}}}} \; = \nonumber \\
& = & \frac{\| \phi_{20}\|^2}{\| \phi_{10}\|^2 + \|
\phi_{20}\|^2},
\end{eqnarray}
which (neglecting the overlapping between the two Gaussian
wavefunctions) corresponds to the {\it standard} quantum
prescription for the probability that $\phi^{\makebox{\tiny D}}$
collapses to $\phi_{2}$.

\noindent 4. The {\it delocalization probability}, i.e. the
probability $P_{\phi_{2}}^{\makebox{\tiny Del}}$ that ${\tilde
\Gamma}_{t}^{\makebox{\tiny R}}$ goes below $b - \eta$ before time
$s_{\infty}$, after having reached $b$ ($\eta$ is a positive
quantity smaller than $b$), is\footnote{See formula 6, page 117 of
\cite{gsb}.}:
\begin{eqnarray}
P_{\phi_{2}}^{\makebox{\tiny Del}} & \leq & 1 \; - \; P\{ \inf
{\tilde
\Gamma}_{s}^{\makebox{\tiny R}} > b - \eta \} \; = \; \nonumber \\
& = & 1 \; - \; (1 + \tanh b)\, \frac{\tanh \eta}{1 + \tanh \eta}.
\end{eqnarray}
In the above definition, we have required a delocalization to
occur before time $s_{\infty}$ because --- since $s_{\infty}$
corresponds to $t = \infty$ --- a delocalization at a time $\geq
s_{\infty}$ does not correspond to a real physical delocalization.

If, for example, we take $\eta = 3$, we have that
$P_{\phi_{2}}^{\makebox{\tiny Del}} \lesssim 0.002$.

This concludes our analysis of the statistical behavior of ${\tilde
\Gamma}_{s}^{\makebox{\tiny R}}$. We end this section by discussing
how one can estimate the error made in neglecting the term
$g_{t}(\Gamma_{t}^{\makebox{\tiny R}})$ in Eq. (\ref{eqfg2}), i.e.
in studying ${\tilde \Gamma}_{s}^{\makebox{\tiny R}}$ in place of
$\Gamma_{s}^{\makebox{\tiny R}}$. In Appendix A the following
estimate for $g_{t}(\Gamma_{t}^{\makebox{\tiny R}})$ is given:
\begin{eqnarray} \label{app2}
\left| g_{t}(\Gamma_{t}^{\makebox{\tiny R}}) \right| & \leq &
\lambda \frac{(|X_{t}| + |Y_{t}|)^2}{e^{a_{t}^{\makebox{\tiny
R}}(|X_{t}| + |Y_{t}|)^2/4} - 1 } \\
& \leq &  \frac{\lambda X_{t}^2}{e^{a_{m}^{\makebox{\tiny R}}\,
X_{m}^2/4} - 1 }, \label{app3}
\end{eqnarray}
where $a_{m}^{\makebox{\tiny R}}$ and $X_{m}$ are the minimum
values $a_{t}^{\makebox{\tiny R}}$ and $X_{t}$ take during the
time interval one is considering. For example, if we take a 1--g
object and we assume that $a_{m}^{\makebox{\tiny R}} =
a_{\infty}^{\makebox{\tiny R}}$ and $X_{m} = 1$ cm, then we have:
\begin{equation}
c \; \equiv \; \frac{1}{e^{a_{m}^{\makebox{\tiny R}}\, X_{m}^2/4}
- 1 } \; \simeq \; e^{-10^{21}},
\end{equation}
which is very close to zero.

By using inequality (\ref{app3}) and lemma 4, pag. 120 of
\cite{gsb}, one can easily show that $\Gamma_{s}^{\makebox{\tiny
R} -} \leq \Gamma_{s}^{\makebox{\tiny R}} \leq
\Gamma_{s}^{\makebox{\tiny R} +}$, where
$\Gamma_{s}^{\makebox{\tiny R} \pm}$ are solutions of the
following two stochastic differential equations:
\begin{equation} \label{eqforpm}
d \Gamma_{s}^{\makebox{\tiny R} \pm} \; = \; \left[ \tanh
{\Gamma}_{s}^{\makebox{\tiny R} \pm} \pm c \right] ds \, + \,
d{\tilde W}_{s},
\end{equation}
with an obvious meaning of the signs (as before, we have moved from
the variable $t$ to the variable $s$).

Eqs. (\ref{eqforpm}) can be analyzed along the same lines followed
in studying Eq. (\ref{eqfg4}), getting basically the same results,
due to the very small value $c$ takes in most relevant physical
situations. This kind of analysis will be done in detail in a future
paper: there, we will study the most important case where a
macroscopic superposition can be created, i.e. a measurement--like
situation in which a macroscopic object acting like a measuring
apparatus interacts with a microscopic system being initially in a
superposition of two eigenstates of the operator which is measured;
we will show that, throughout the interaction, the wavefunction of
the apparatus is
--- with extremely high probability --- always localized in space,
and that the measurement has a definite outcome with the correct
quantum probabilities.

\section{General solution: the asymptotic behavior} \label{gs}

In this section we analyze the asymptotic behavior of the general
solution of Eq. (\ref{nle}), showing that --- as time increases ---
any wavefunction collapses towards a stationary Gaussian solution.

\subsection{The collapse}

We have seen in the previous sections that both the
single--Gaussian and double--Gaussian solutions asymptotically
converge towards a stationary solution having the form:
\begin{equation} \label{asol}
\psi_{t}(x) \; = \; \sqrt[4]{\frac{2a^{\makebox{\tiny
R}}_{\infty}}{\pi}}\, \exp\left[ -a_{\infty}(x -
\overline{x}_{t})^2 + i \overline{k}_{t} x \right],
\end{equation}
with:
\begin{equation}
a_{\infty} \; = \; \lim_{t \rightarrow +\infty} a_{t} \; = \;
\frac{\lambda}{\omega} (1-i);
\end{equation}
it becomes then natural to ask whether such a kind of wavefunction
is the asymptotic limit of {\it any} initial wavefunction. The
answer is positive \cite{hal} as we shall now see by following the
same strategy used in refs. \cite{jp2,ad2,hal} to prove
convergence of solutions.

Since a wavefunction of the form (\ref{asol}) is an eigenstate of
the operator:
\begin{equation}
A \; = \; q \, + \, \frac{i - 1}{m\omega} \, p,
\end{equation}
the proof consists in showing that the variance\footnote{In fact,
it is easy to prove that $\Delta A_{t} = 0$ if and only if $|
\psi_{t} \rangle$ is an eigenstate of the operator $A$, from which
it follows that $\Delta A_{t} \rightarrow 0$ if and only if $|
\psi_{t} \rangle$ converges towards an eigenstate of $A$. In Eq.
(\ref{var}), $\langle A \rangle \equiv \langle \psi_{t}| A
\psi_{t}\rangle$, and similarly for $\langle A^{\dagger}
\rangle$.}:
\begin{equation} \label{var}
\Delta A_{t} \; \equiv \; \langle \psi_{t} |[ A^{\dagger} -
\langle A^{\dagger} \rangle] [A - \langle A \rangle] | \psi_{t}
\rangle
\end{equation}
converges to 0 for $t \rightarrow +\infty$. The following
expression for $\Delta A_{t}$ holds:
\begin{equation}
\Delta A_{t} \; = \; \Delta q_{t} + \frac{2}{m^2 \omega^2} \Delta
p_{t} - \frac{2}{m \omega} \Sigma(q,p) - \frac{\hbar}{m \omega},
\end{equation}
with:
\begin{eqnarray}
\Sigma(q,p) & = & \frac{1}{2}\left[ \langle \psi_{t} |[ q -
\langle q \rangle] [p - \langle p \rangle] | \psi_{t} \rangle +
\right.
\nonumber \\
& & \;\;\;\; \left. \langle \psi_{t} |[ p - \langle p \rangle] [q
- \langle q \rangle] | \psi_{t} \rangle \right]
\end{eqnarray}
After a rather long calculation, one finds that:
\begin{eqnarray} \label{eqvar}
\frac{d}{dt}\, {\mathbb E} \left[ \Delta A_{t} \right] & = & -
{\mathbb E} \left[ \omega\, \Delta A_{t} + 2\lambda\, \left(
\Delta q_{t} - \sigma_{q}^{2}(\infty) \right)^2
\right. \nonumber \\
& & \left. + 2\lambda\, \left( \Delta q_{t} - \frac{2}{m \omega}\,
\Sigma(q,p) \right)^2 \right] \;
\leq \; 0. \nonumber \\
& &
\end{eqnarray}
Since, by definition, ${\mathbb E} \left[ \Delta A_{t} \right]$ is
a non--negative quantity, Eq. (\ref{eqvar}) is consistent if and
only if the right--hand--side asymptotically vanishes. This, in
particular, implies that
\begin{equation}
\Delta A_{t} \; \xrightarrow[t \,\rightarrow \, +\infty]{} \; 0,
\end{equation}
except for a subset of $\Omega$ of measure zero.

\subsection{The collapse probability}

We now analyze the probability that the wavefunction, as a result of
the collapse process, lies within a given region of
space\footnotemark. To this purpose, let us consider the following
probability measure, defined on the Borel sigma--algebra ${\mathcal
B}({\mathbb R})$:
\begin{equation} \label{pmes}
\mu_{t}(\Delta) \; \equiv \; {\mathbb E}_{{\mathbb P}}\left[
\|P_{\Delta}\psi_{t}\|^2\right],
\end{equation}
where $P_{\Delta}$ is the projection operator associated to the
Borel subset $\Delta$ of ${\mathbb R}$. Such a measure is
identified by the density $p_{t}(x) \equiv {\mathbb E}_{{\mathbb
P}} [| \psi_{t}(x)|^2 ]$:
\begin{equation}
\mu_{t}(\Delta) \; = \; \int_{\Delta} p_{t}(x) \, dx,
\end{equation}
and it can be easily shown that the density $p_{t}(x)$ corresponds
to the diagonal element $\langle x | \rho_{t} | x \rangle$ of the
statistical operator\footnotemark $\rho_{t} \equiv {\mathbb
E}_{{\mathbb P}}\left[ |\psi_{t}\rangle\langle\psi_{t}|\right]$,
which satisfies the Lindblad--type equation:
\begin{equation} \label{lin}
\frac{d}{dt}\, \rho_{t} \; = \; -\frac{i}{2m\hbar}\, \left[ p^2,
\rho_{t} \right] - \frac{\lambda}{2} \left[ q, \left[ q, \rho_{t}
\right] \right].
\end{equation}
The solution of the above equation, expressed in terms of the
solution $\langle q_{1}| \rho_{S}(t) | q_{2} \rangle$ of the pure
Schr\"odinger equation ($\lambda = 0$), is\footnotemark:
\begin{widetext}
\begin{equation} \label{sol_eq}
\langle q_{1}| \rho(t) | q_{2} \rangle \; = \; \frac{1}{2 \pi \hbar}
\int_{-\infty}^{+\infty} dk \int_{-\infty}^{+\infty} dy\,
e^{-(i/\hbar)k y} F[k, q_{1} - q_{2}, t] \langle q_{1} + y|
\rho_{S}(t) | q_{2} + y \rangle,
\end{equation}
with:
\begin{equation}
F[k,x,t] \; = \; \exp\left[ -\frac{\lambda}{2}\, t\, \left( x^2 -
\frac{k}{m}\, x t + \frac{k^2}{3 m^2} t^2 \right) \right].
\end{equation}
\end{widetext}
The Hermitian symmetry of $\langle q_{1}| \rho(t) | q_{2} \rangle$
follows from the fact that $F[k,x,t] = F[-k, -x, t]$; for $\lambda =
0$ we have $F[k,x,t] = 1$ so that $\langle q_{1}| \rho(t) | q_{2}
\rangle = \langle q_{1}| \rho_{S}(t) | q_{2} \rangle$ as it must be.

From Eq.~\eqref{sol_eq} it follows that:
\begin{equation}
p_{t}(x) = \frac{1}{2 \pi \hbar} \int_{-\infty}^{+\infty}
dk\int_{-\infty}^{+\infty} dy\, e^{-(i/\hbar)ky} F[k, 0, t]
p_{t}^{S}(x + y)
\end{equation}
where $p_{t}^{S}(x) = \langle x| \rho_{S}(t) | x \rangle$ is the
standard quantum probability density of finding the particle located
$x$ in a position measurement. One can easily perform the
integration over the $k$ variable and he gets:
\begin{equation} \label{fin_form_prob}
p_{t}(x) \; = \; \sqrt{\frac{\alpha_{t}}{\pi}}
\int_{-\infty}^{+\infty} dy\, e^{- \alpha_{t} y^2} p_{t}^{S}(x + y),
\end{equation}
with:
\begin{equation}
\alpha_{t} \; = \; \frac{3 m_{0}}{2 \hbar^2 \lambda_{0}} \,
\frac{m}{t^3} \; \simeq \; 10^{43} \left( \frac{m}{\makebox{kg}}
\right) \left( \frac{\makebox{sec}}{t} \right)^3.
\end{equation}
\addtocounter{footnote}{-2} \footnotetext{Of course, wavefunctions
in general do not have a compact support; accordingly, saying that a
wavefunction lies within a given region of space amounts to saying
that it is almost entirely contained within the region, except for
small ``tails'' spreading out of that region.}For a macroscopic
object (let us say $m \geq 1$ g) and for very long times (e.g. $t
\in [0, 10^8$ sec $\simeq$ 3 years], a time interval which of course
is much longer than the time during which an object can be kept
isolated so that the free particle approximation holds true)
$\alpha_{t}$ is a very large number, so large that the exponential
function in Eq.~\eqref{fin_form_prob} is significantly more narrow
than $p_{t}^{S}(x)$, for most typical wavefunctions (in
Sec.~\ref{amp-mech} we will see that the asymptotic spread of the
wavefunction of a 1--g object is about $10^{-13}$ m). Accordingly,
this exponential function acts like a Dirac--delta and one has:
$p_{t}(x) \simeq p^{S}_{t}(x)$, with very high accuracy. This in
turn implies that:
\begin{equation}
\mu_{t}(\Delta) \; \simeq \; \|P_{\Delta}\psi^{\makebox{\tiny
Sch}}_{t}\|^2:
\end{equation}
in other words, the probability measure $\mu_{t}(\Delta)$ is very
close to the {\it quantum probability} of finding the particle lying
in $\Delta$ in a position measurement.

\stepcounter{footnote}\footnotetext{The definition $\rho_{t} \equiv
{\mathbb E}_{{\mathbb P}}\left[
|\psi_{t}\rangle\langle\psi_{t}|\right]$ is rather formal; see
\cite{bar3,hol1} for a rigorous definition.}

\stepcounter{footnote}\footnotetext{Ref. \cite{grw1}, Appendix C,
shows how the solution can be obtained.}

We still have to discuss the physical meaning of the probability
measure $\mu_{t}(\Delta)$ defined by Eq. (\ref{pmes}); such a
discussion is relevant only at the {\it macroscopic} level, since we
need only macro--objects to be well localized in space (and with the
correct quantum probabilities).

As already anticipated the spread of a wavefunction of a
macro--object having the mass of e.g. 1 g reaches in a very short
time a value which is close to the asymptotic spread
$\sigma_q(\infty) \simeq 10^{-13}$ m; then, if we take for $\Delta$
an interval $[a, b]$ whose length is much greater than
$\sigma_q(\infty)$ --- for example, we can take $b - a = 10^{-7}$ m,
which is sufficiently small for all practical purposes
--- only those wavefunctions whose mean lies around $\Delta$ give a
non vanishing contribution to $\mu_{t}(\Delta)$. As a consequence,
with the above choices for $\Delta$ (and, of course, waiting a time
sufficiently long in order for the reduction to have occurred), the
measure $\mu_{t}(\Delta)$ represents a good probability measure that
the wavefunction collapses within $\Delta$.

\section{Effect of the reducing terms on the microscopic dynamics}
\label{micro}

In the previous sections we have studied some analytical
properties of the solutions of the stochastic differential
equation (\ref{nle}); we now focus our attention on the dynamics
for a microscopic particle.

The time evolution of a (free) quantum particle has three
characteristic features:

\noindent 1. The wavefunction is subject to a localization process,
which, at the micro--level, is extremely slow, almost negligible.
For example, with reference to the situation ${\mathcal A}$ we have
defined at the end of Sec. \ref{twogaussas}, Eq. (\ref{sita}) shows
that the distance $X_{t}$ between the centers of the two Gaussian
wavefunctions remains practically unaltered for about $10^5$ sec;
under the approximation $X_t \simeq X_0$, Eqs. (\ref{tch}) and
(\ref{ats}) imply that the time necessary for one of the two
Gaussians to be suppressed is:
\begin{eqnarray}
{\mathbb E}\left[ T_{b} \right] & \simeq & 10^6\, \left(
\frac{\makebox{m}}{X_0} \right)^{2} \;
\makebox{sec $\quad$ for an electron,} \nonumber \\
{\mathbb E}\left[ T_{b} \right] & \simeq & 10^3\, \left(
\frac{\makebox{m}}{X_0} \right)^{2} \; \makebox{sec $\quad$ for a
nucleon,} \nonumber
\end{eqnarray}
which are very long times\footnote{In the first case, indeed, the
reduction time is longer than the time during which the
approximation $X_t \simeq X_0$ is valid --- see Eq.
(\ref{sita}).}, compared with the characteristic times of a
quantum experiment.

\noindent 2. The spread of the wavefunction reaches an asymptotic
value which depends on the mass of the particle:
\begin{eqnarray}
\sigma_{q}(\infty) & \simeq & 1 \;
\makebox{m $\quad\;\,$ for an electron,} \nonumber \\
\sigma_{q}(\infty) & \simeq & 1 \; \makebox{cm $\quad$ for a
nucleon.} \nonumber
\end{eqnarray}

\noindent 3. The wavefunction undergoes a random motion both in
position and momentum, under the influence of the stochastic
process. These fluctuations can be quite relevant at the
microscopic level; for example, for a stationary solution, one has
from Eq. (\ref{eq1c}):
\begin{eqnarray}
\frac{d}{dt}\,{\mathbb V}\left[ \langle q \rangle_{t} \right] &
\geq & 1 \quad\! \makebox{m$^2$/sec $\qquad$ for an electron,}
\nonumber  \\
\frac{d}{dt}\,{\mathbb V}\left[ \langle q \rangle_{t} \right] &
\geq & 10^{-3} \; \makebox{m$^2$/sec $\quad$ for a nucleon.}
\nonumber
\end{eqnarray}

This is the physical picture of a microscopic particle as it
emerges from the collapse model. In order for the model to be
physically consistent, it must reproduce at the microscopic level
the predictions of standard quantum mechanics, an issue which we
are now going to discuss.

Within the collapse model, measurable quantities are given
\cite{mis} by averages of the form ${\mathbb E}_{{\mathbb
P}}[\langle O \rangle_{t}]$, where $O$ is (in principle) any
self--adjoint operator. It is not difficult to prove that:
\begin{equation}
{\mathbb E}_{{\mathbb P}}\left[\langle O \rangle_{t}\right] \; =
\; \makebox{Tr} \left[ O \rho_{t} \right],
\end{equation}
where the statistical operator $\rho_{t} \equiv {\mathbb
E}_{{\mathbb P}}\left[ |\psi_{t}\rangle\langle\psi_{t}|\right]$
satisfies the Lindblad--type Eq.~\eqref{lin}. This is a typical
master equation used in decoherence theory to describe the
interaction between a quantum particle and the surrounding
environment \cite{dec1}; consequently, as far as {\it experimental
results} are concerned, the predictions of our model are similar to
those of decoherence models\footnote{We recall the conceptual
difference between collapse models and decoherence models. Within
collapse models, one modifies quantum mechanics (by adding
appropriate nonlinear and stochastic terms) so that macro--objects
are always localized in space. Decoherence models, on the other
hand, are quantum mechanical models applied to the study of open
quantum system; since they assume the validity of the Schr\"odinger
equation, they cannot induce the collapse of the wavefunction of
macroscopic systems (as it has been shown, e.g. in \cite{bg}) even
if one of the effects of the interaction with the environment is to
{\it hide} --- not to {\it eliminate} --- macroscopic superpositions
in measurement--like situations.}. It becomes then natural to
compare the strength of the collapse mechanism (measured by the
parameter $\lambda$) with that of decoherence.

Such a comparison is given in Table \ref{tab1}, when the system
under study is a very small particle like an electron, or an
almost macroscopic object like a dust particle.
\begin{table}
\begin{tabular}{||l|cc||} \hline
Cause of decoherence & $\quad$ $10^{-3}$ cm  $\quad$ & $\quad$ $10^{-6}$ cm $\quad$ \\
 & dust particle  & large molecule \\ \hline
Air molecules  & $10^{36}$ & $10^{30}$  \\
Laboratory vacuum  & $10^{23}$ & $10^{17}$ \\
Sunlight on earth  & $10^{21}$ & $10^{13}$ \\
300K photons  & $10^{19}$ & $10^{6}$ \\
Cosmic background rad. & $10^{6}$ & $10^{-12}$ \\
\hline \hline
COLLAPSE  & $10^{7}$ & $10^{-2}$ \\
\hline
\end{tabular}
\caption{$\lambda$ in cm${}^{-2}$sec${}^{-1}$ for decoherence
arising from different kinds of scattering processes (taken from
Joos and Zeh \cite{dec1}). In the last line: $\lambda$ for the
collapse model given by Eq. (\ref{nle}).}\label{tab1}
\end{table}
We see that, for most sources of decoherence, the {\it
experimentally testable} effects of the collapse mechanism are
weaker than those produced by the interaction with the surrounding
environment. This implies that, in order to test such effects, one
has to isolate a quantum system --- for a sufficiently long time
--- from almost all conceivable sources of decoherence, which is
quite difficult: the experimentally testable differences between our
collapse model and standard quantum mechanics are so small that they
cannot be detected unless very sophisticated experiments are
performed \cite{pref,exp}.

\section{Multi--particle systems: effect of the reducing terms
on the macroscopic dynamics} \label{macro}

The generalization of Eq. (\ref{nle}) to a system of $N$
interacting and distinguishable particles is straightforward:
\begin{eqnarray} \label{nlemp}
d\,\psi_{t}(\{ x \}) & = & \left[ -\frac{i}{\hbar}\,
H_{\makebox{\tiny T}}\, dt + \sum_{n=1}^{N}\sqrt{\lambda_{n}}\, (
q _{n} - \langle q _{n} \rangle_{t})\, d W_{t}^{n}\right. \nonumber \\
& & \left. - \frac{1}{2}\,\sum_{n=1}^{N} \lambda_{n} ( q _{n} -
\langle  q _{n} \rangle_{t})^2 dt \right] \psi_{t}(\{ x \}),
\end{eqnarray}
where $H_{\makebox{\tiny T}}$ is the standard quantum Hamiltonian
of the composite system, the operators $ q _{n}$ ($n = 1, \ldots
N$) are the position operators of the particles of the system, and
$W_{t}^{n}$ ($n = 1, \ldots N$) are $N$ independent standard
Wiener processes; the symbol $\{ x \}$ denotes the $N$ spatial
coordinates $x_{1}, \ldots x_{N}$.

For the purposes of our analysis, it is convenient to switch to
the center--of--mass ($R$) and relative ($\tilde{x}_{1},
\tilde{x}_{2}, \ldots \tilde{x}_{N}$) coordinates:
\begin{equation}
\left\{
\begin{array}{lcl}
R & = & \displaystyle \frac{1}{M}\, \sum_{n=1}^{N}
m_{n}\, x_{n} \qquad M \; = \; \sum_{n=1}^{N} m_{n},\\
& & \\
x_{n} & = & R + \tilde{x}_{n};
\end{array}
\right.
\end{equation}
let $Q$ be the position operator for the center of mass and
$\tilde q _{n}$ ($n = 1 \ldots N$) the position operators
associated to the relative coordinates.

It is not difficult to show that --- under the assumption
$H_{\makebox{\tiny T}} = H_{\makebox{\tiny CM}} +
H_{\makebox{\tiny rel}}$ --- the dynamics for the center of mass
and that for the relative motion decouple; in other words,
$\psi_{t}(\{ x \}) = \psi_{t}^{\makebox{\tiny CM}}(R) \otimes
\psi_{t}^{\makebox{\tiny rel}}(\{\tilde{x}\})$ solves Eq.
(\ref{nlemp}) whenever $\psi_{t}^{\makebox{\tiny CM}}(R)$ and
$\psi_{t}^{\makebox{\tiny rel}}(\{\tilde{x}\})$ satisfy the
following equations:
\begin{eqnarray}
d\,\psi_{t}^{\makebox{\tiny rel}}(\{\tilde{x}\}) & = & \left[
-\frac{i}{\hbar}\, H_{\makebox{\tiny rel}}\, dt +
\sum_{n=1}^{N}\sqrt{\lambda_{n}}\, ({\tilde{q}}_{n} - \langle
{\tilde{q}}_{n} \rangle_{t}) d W_{t}^{n}\right. \nonumber \\
& & \left. - \frac{1}{2}\,\sum_{n=1}^{N} \lambda_{n}
({\tilde{q}}_{n} - \langle {\tilde{q}}_{n} \rangle_{t})^2 dt
\right] \psi_{t}^{\makebox{\tiny rel}}(\{ x \}), \\ & & \nonumber \\
d\,\psi_{t}^{\makebox{\tiny CM}}(R) & = & \left[
-\frac{i}{\hbar}\, H_{\makebox{\tiny CM}}\, dt +
\sqrt{\lambda_{\makebox{\tiny CM}}}\, (Q - \langle
Q \rangle_{t}) d W_{t}\right. \nonumber \\
& & \left. - \frac{\lambda_{\makebox{\tiny CM}}}{2}\, (Q - \langle
Q \rangle_{t})^2 dt \right] \psi_{t}^{\makebox{\tiny CM}}(R),
\label{eqcm}
\end{eqnarray}
with:
\begin{equation}
\lambda_{\makebox{\tiny CM}} \; = \; \sum_{n=1}^{N} \lambda_{n} \;
= \; \frac{M}{m_{0}}\, \lambda_{0}.
\end{equation}
The first of the above equations describes the internal motion of
the composite system, and will not be analyzed in this paper; in the
remainder of the section, we will focus our attention on the second
equation.

Eq. (\ref{eqcm}) shows that the reducing terms associated to the
center of mass of a composite system are equal to those associated
to a particle having mass equal to the total mass $M$ of the
composite system; in particular, when the system is isolated ---
i.e., $H_{\makebox{\tiny CM}} = P^2/2M$, where $P$ is the total
momentum --- the center of mass behaves like a free particle, whose
dynamics has been already analyzed in Sects. \ref{1gauss},
\ref{2gauss} and \ref{gs}. In the next subsections we will see that,
because of the large mass of a macro--object, the dynamics of its
center of mass is radically different from that of a microscopic
particle.

\subsection{The amplification mechanism}
\label{amp-mech}

The first important feature of collapse models is what has been
called the ``amplification mechanism'' \cite{grw1,grw2}: the
reduction rates of the constituents of a macro--object sum up, so
that the reduction rate associated to its center of mass is much
greater than the reduction rates of the single constituents.

\begin{figure}
\begin{center}
{\includegraphics[scale=0.5]{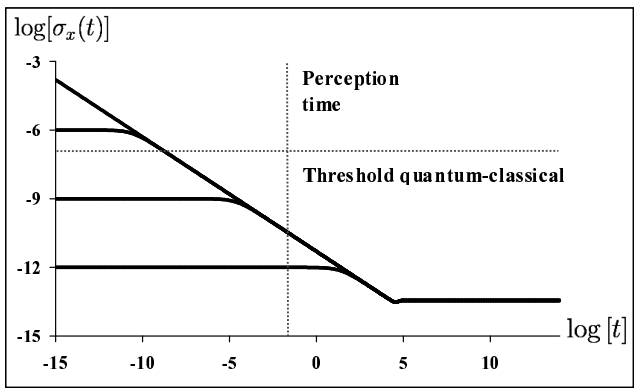}} \caption{Time evolution
of the spread $\sigma_{q}(t)$ of a Gaussian wavefunction of the
center of mass of a system containing $N = 10^{24}$ nucleons;
different initial conditions have been considered. Time is measured
in seconds, while the spread is measured in meters.} \label{fig5}
\end{center}
\end{figure}

This situation is exemplified in Fig. \ref{fig5}, which shows the
time evolution of the spread $\sigma_{q}(t)$ of a Gaussian
wavefunction.
We note that, however large the initial wavefunction is, after
less than $10^{-2}$ sec --- which corresponds to the perception
time of a human being --- its spread goes below $10^{-5}$ cm,
which is the threshold chosen in the original GRW model
\cite{grw1}, below which a wavefunction can be considered
sufficiently well localized to describe the classical behavior of
a macroscopic system.

More generally, Eq. (\ref{eqvar}) implies that:
\begin{equation}
\left| \frac{d}{dt}\, {\mathbb E} \left[ \Delta A_{t} \right]
\right| \; \geq \; 2\lambda \left[ \Delta q \, - \,
\sigma^{2}_{q}(\infty) \right]^2,
\end{equation}
and, as long as the spread of the wavefunction is significantly
greater than its asymptotic value, i.e. the wavefunction is not
already sufficiently well localized in space, we have:
\begin{equation}
\left| \frac{d}{dt}\, {\mathbb E} \left[ \Delta A_{t} \right]
\right| \; \geq \; 10^{25}\, \left( \frac{m}{\makebox{Kg}} \right)
\left(\frac{\Delta q}{\makebox{m}}\right)^2 \;\;\;
\makebox{m$^2$/sec},
\end{equation}
which is a very high reduction rate, for a macroscopic object.
Note that, as previously stated, the velocity increases, for
increasing values of the mass of the particle.

Asymptotically, the wavefunction of a macro--object has an
extremely small spread; for example:
\begin{eqnarray}
\sigma_{q}(\infty) & \simeq & 10^{-13} \;
\makebox{m $\quad$ for an 1--g object,} \nonumber \\
\sigma_{q}(\infty) & \simeq & 10^{-27} \; \makebox{m $\quad$ for
the Earth.} \nonumber
\end{eqnarray}
Thus, according to our collapse model, macro--particles behave
like point--like particles.

\subsection{Damping of fluctuations}

We have seen that the mean (both in position and in momentum) of the
wavefunction undergoes a diffusion process arising from the
stochastic dynamics: such a diffusion is quite important at the
microscopic level, and it is responsible for the agreement of the
physical predictions of the model with those given by standard
quantum mechanics. We now analyze the magnitude of the fluctuations
at the macroscopic level.

Contrary to the behavior of the reduction mechanism, which is
amplified when moving from the micro-- to the macro--level, the
fluctuations associated to the motion of microscopic particles
interfere destructively with each other, in such a way that the
diffusion process associated to the center of mass of an
$N$--particle system is much weaker than those of the single
components. We now give some estimates.

Let us suppose that the center--of--mass wavefunction has reached
a stationary solution; under this assumption, one has from Eqs.
(\ref{eq1c}), (\ref{eq2c}) and (\ref{eq3c}):
\begin{eqnarray}
{\mathbb V}[\langle q \rangle_{t}] & = & \frac{\omega}{8 \lambda}
\left[ \frac{(\omega t)^3}{3!}\, + \, \frac{(\omega t)^2}{2!}\, +
\, \omega t \right],\\
{\mathbb V}[\langle p \rangle_{t}] & = & \lambda \hbar^2 t.
\end{eqnarray}
Since, for example, $\omega/8\lambda \simeq 10^{-27}$ m$^2$ for a
1--g object, and $\omega/8\lambda \simeq 10^{-54}$ m$^2$ for the
Earth, we see that for a macro--object the numerical values of the
parameters are such that for very long times (in many cases much
longer that the age of the universe) the fluctuations are so small
that, for all practical purposes, they can be neglected; this is
how classical determinism is recovered within our stochastic
model.

The above results imply that the actual values of $\langle q
\rangle_{t}$ and $\langle p \rangle_{t}$ are practically
equivalent to their stochastic averages; since these stochastic
averages obey the classical laws of motion (\ref{mpa}) and
(\ref{emim}), we find out that $\langle q \rangle_{t}$ and
$\langle p \rangle_{t}$ practically evolve according to the
classical laws of motion, for most realizations of the stochastic
process.

The conclusion is the following: in the macroscopic regime, the
wavefunction of a macroscopic system behaves, for all practical
purposes, like a point--like particle moving deterministically
according to Newton laws of motion.

\section{Conclusions} \label{concl}

From the analysis of the previous sections we have seen that, in
general, the evolution of the wavefunction as predicted by the
collapse model is significantly different from that predicted by
standard quantum mechanics, both at the micro-- and at the
macro--level. For example, at the microscopic level the random
fluctuations can be very large, while in the standard case there are
no fluctuations; at the macroscopic level, wavefunctions rapidly
localize in space, while in the standard quantum case they keep
spreading.

Anyway, as far as {\it physical predictions} are concerned, our
model is almost equivalent to standard quantum mechanics, the
differences being so small that they can hardly be detected with
present--day technology. Moreover, at the macroscopic level the
localization mechanism becomes very rapid and the fluctuations
almost disappear: the wavefunction of the center of mass of a
macroscopic object behaves like a point--like particle moving
according to Newton's laws.

To conclude, the stochastic model reproduces, with excellent
accuracy, both quantum mechanics at the microscopic level and
classical mechanics at the macroscopic one, and describes also the
transition from the quantum to the classical domain.

\section*{Acknowledgements}

We acknowledge very stimulating discussions with S.L. Adler, D.
D\"urr, G.C. Ghirardi, E. Ippoliti, P. Pearle, D.G.M. Salvetti and
B. Vacchini.

\appendix
\section{Derivation of $\;\;\;$ inequality (\ref{app2})}

From Eq. (\ref{efg}), one has:
\begin{eqnarray}
\lefteqn{\left| g_{t}(\Gamma_{t}^{\makebox{\tiny R}}) \right| \;
\leq} & & \\
& & 2 \lambda |X_{t}| h_t \left[ |Y_{t} \sin \theta_{t}| \, \left|
\frac{e^{\Gamma_{t}^{\makebox{\tiny R}}}}{1 +
e^{2\Gamma_{t}^{\makebox{\tiny R}}} + 2 h_t \cos \theta_{t}
e^{\Gamma_{t}^{\makebox{\tiny R}}}} \right| + \right.
\nonumber \\
& & \left. | X_{t} \cos \theta_{t} | \left|
\frac{e^{\Gamma_{t}^{\makebox{\tiny R}}}}{1 +
e^{2\Gamma_{t}^{\makebox{\tiny R}}} + 2 h_t \cos \theta_{t}
e^{\Gamma_{t}^{\makebox{\tiny R}}}} \right| \left| \frac{1 -
e^{2\Gamma_{t}^{\makebox{\tiny R}}}}{1 +
e^{2\Gamma_{t}^{\makebox{\tiny R}}}} \right| \right]. \nonumber
\end{eqnarray}
By using the fact that $\sin \theta_{t} \leq 1$ and $\cos
\theta_{t} \leq 1$, that the function
\begin{equation}
f_{1}(x) \; = \; \frac{x}{1 + x^2 + 2cx} \qquad x \geq 0, \quad -1
< c \leq 1
\end{equation}
is bounded between 0 and $1/(2 + 2c)$, while the function
\begin{equation}
f_{2}(x) \; = \; \frac{1 - x}{1 + x} \qquad x \geq 0
\end{equation}
is bounded between 1 and $-1$, we get:
\begin{eqnarray}
\left| g_{t}(\Gamma_{t}^{\makebox{\tiny R}}) \right| & \leq &
\lambda \, \frac{h_{t}}{1 - h_{t}}\, \left[ X_{t}^2 + | X_{t}
Y_{t} | \right] \\
& \leq & \lambda \, \frac{h_{t}}{1 - h_{t}}\, \left[ |X_{t}| +
|Y_{t}| \right]^2, \nonumber
\end{eqnarray}
from which, dividing both the numerator and the denumerator on the
right--hand--side by $h_{t}$ and using $(x + y)^2 \leq 2(x^2 +
y^2)$, inequality (\ref{app2}) follows. \\

\section*{Note (first published in: A. Bassi, D.G.M. Salvetti, J.
Phys. A: 40, 9859 (2007))}

Eqs.~\eqref{p4} and ~\eqref{p5} contain a computational mistake. The
correct equations are:
\begin{eqnarray}
d\gamma_{t}^{\makebox{\tiny R}} & = & \left[ \lambda \bar{x}_{t}^{2}
+ \frac{\hbar}{m}\, a^{\makebox{\tiny I}}_{t} +
\frac{\lambda}{4 a^{\makebox{\tiny R}}_{t}}\right]dt \qquad\qquad\qquad\qquad\nonumber \\
& & + \; \sqrt{\lambda}\, \bar{x}_{t} \left[ d\xi_{t} -
2\sqrt{\lambda}\, \bar{x}_{t}\, dt
\right] \label{eq:ap1} \\
d\gamma_{t}^{\makebox{\tiny I}} & = & \left[ - \frac{\hbar}{m}\,
a^{\makebox{\tiny R}}_{t}  - \frac{\hbar}{2m}\, \bar{k}_{t}^{2}+
\frac{\lambda a^{\makebox{\tiny I}}_{t}}{4 (a^{\makebox{\tiny
R}}_{t})^2} \right]dt \nonumber \\
& & +  \sqrt{\lambda} \frac{a^{\makebox{\tiny
I}}_{t}}{a^{\makebox{\tiny R}}_{t}}\,\bar{x}_{t} \left[ d\xi_{t} -
2\sqrt{\lambda}\, \bar{x}_{t}\, dt \right], \label{eq:ap2}
\end{eqnarray}
which differ from Eqs.~\eqref{p4} and ~\eqref{p5}, in the first case
for the extra factor $\lambda/4 a^{\makebox{\tiny R}}_{t}$ and in
the second case for the factor $\lambda a^{\makebox{\tiny I}}_{t}/4
(a^{\makebox{\tiny R}}_{t})^2$. We correct in this way a mistake,
which however does not affect the subsequent analysis.

\end{document}